\documentclass[fleqn,twoside]{article}
\usepackage{espcrc2}

\usepackage{xspace}
\usepackage{url}
\usepackage{graphicx}
\usepackage[figuresright]{rotating}

\newcommand{\AmS}{{\protect\the\textfont2
  A\kern-.1667em\lower.5ex\hbox{M}\kern-.125emS}}

\newcommand{\mev}{\ensuremath{\mathrm{Me\kern -0.1em V}}\xspace}
\newcommand{\gev}{\ensuremath{\mathrm{\,Ge\kern -0.1em V}}\xspace}
\def\Vus  {\ensuremath{|V_{us}|}\xspace}
\def\Vud  {\ensuremath{|V_{ud}|}\xspace}
\def\mtau       {\ensuremath{\tau}\xspace}

\title{Status Report from Tau subgroup of the HFAG}

\author{Swagato Banerjee\address[UVIC]{University of Victoria, Canada.},
        Kiyoshi Hayasaka\address{Nagoya University, Japan.},
        Hisaki Hayashii\address{Nara Womana's University, Japan.},
        Alberto Lusiani\address{Scuola Normale Superiore and INFN Pisa, Italy.},
        J.~Michael Roney\addressmark[UVIC],
        Boris Shwartz\address{Budker Institute of Nuclear Physics, Russia.}}
      
\begin{document}

\begin{abstract}
The aim of Tau subgroup of the HFAG is to provide average values
of the mass and branching fractions of the tau lepton.
Using the branching fractions, we present tests of charged current
lepton universality and obtain estimates for $|V_{us}|$.
We also summarize the status of searches for lepton flavor violating
decays of the tau lepton.
\vspace{1pc}
\end{abstract}

\maketitle

The latest averages from the HFAG have been published in Ref.~\cite{HFAG:2010qj}.
Online updates of the Tau Section are available at \url{http://www.slac.stanford.edu/xorg/hfag/tau/}.

\section{Mass of the $\tau$ lepton}
\label{sec:Tau_Mass}

\begin{figure}[!hbtp]
\begin{center}
\includegraphics[height=.42\textheight,width=.49\textwidth]{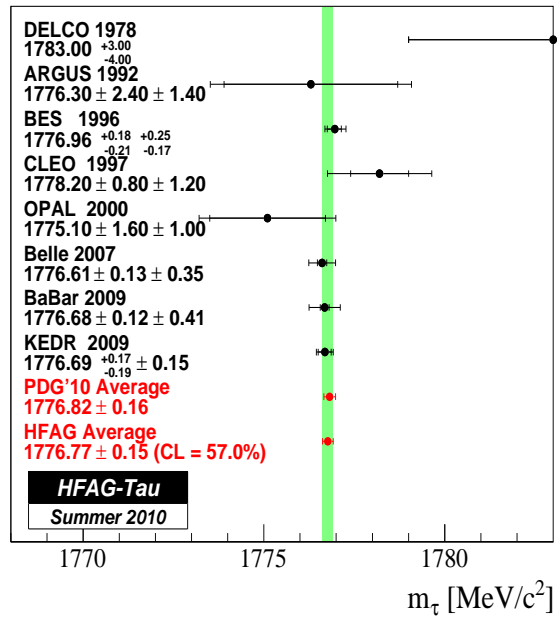}
\end{center}
\caption{Measurements and average value of $m_\tau$.}
\label{fig:Tau_Mass}
\end{figure}

The mass of the $\tau$ lepton has recently been measured by the BaBar~\cite{Aubert:2009ra} and Belle~\cite{Abe:2006vf} experiments
using the end-point technique from the pseudo-mass distribution in $\tau^-\to\pi^-\pi^-\pi^+\nu_\tau$ decays,
as well as by the KEDR experiment~\cite{Shamov:2009zz} from a study of the $\tau^+\tau^-$ cross-section around the production threshold.
In Figure~\ref{fig:Tau_Mass}, we present the measurements and average values of $m_\tau$.

\section{$\tau$ Branching Fractions:}
\label{sec:Tau_BR}

Average values of the $\tau$ branching fractions~\footnote{charge conjugate $\tau$ decays are implied throughout}
are obtained from a global unitarity constrained fit.
We account for correlations due to common dependence of
normalization on the $\tau-$pair cross-section~\cite{Banerjee:2007is}
and assumed branching fractions for the background modes. 
The detector-specific systematics uncertainties are considered to be fully correlated between
measurements of the same experiment.

We use 131 measurements from non-B-factory experiments, 
which includes the set used in the global fit performed by the PDG~\cite{PDG_2010}. 
The measurements from non-B-factories include
     37 measurements from ALEPH,
      2 measurements from ARGUS,
      1 measurement from CELLO,
     36 measurements from CLEO,
      6 measurements from CLEO3,
     14 measurements from DELPHI,
      2 measurements from HRS,
     11 measurements from L3,
     19 measurements from OPAL, and
      3 measurements from TPC.
Finally, we include the following measurements from the B-factories (BaBar and Belle):
\begin{itemize}
\item 12 measurements from BaBar:
\begin{eqnarray*}
{{\cal{B}}(\tau^- \to \mu^- \bar{\nu}_\mu \nu_\tau)}/{{\cal{B}}(\tau^- \to e^- \bar{\nu}_e \nu_\tau)}~\cite{Aubert:2009qj},\\
{{\cal{B}}(\tau^- \to \pi^- \nu_\tau)}/{{\cal{B}}(\tau^- \to e^- \bar{\nu}_e \nu_\tau)}~\cite{Aubert:2009qj},\\
{{\cal{B}}(\tau^- \to  K^-  \nu_\tau)}/{{\cal{B}}(\tau^- \to e^- \bar{\nu}_e \nu_\tau)}~\cite{Aubert:2009qj},\\
{\cal{B}}(\tau^- \to K^- \pi^0 \nu_\tau)~\cite{Aubert:2007jh}\\
{\cal{B}}(\tau^- \to \bar{K}^0 \pi^- \nu_\tau)~\cite{Aubert:2008an}\\
{\cal{B}}(\tau^- \to \bar{K}^0 \pi^- \pi^0 \nu_\tau)~\cite{Paramesvaran:2009ec}\\
{\cal{B}}(\tau^- \to \pi^- \pi^- \pi^+ \nu_\tau ~(\mathrm{ex.~}K^0))~\cite{Aubert:2007mh}\\
{\cal{B}}(\tau^- \to  K^-  \pi^- \pi^+ \nu_\tau ~(\mathrm{ex.~}K^0))~\cite{Aubert:2007mh}\\
{\cal{B}}(\tau^- \to  K^-  \pi^-  K^+ \nu_\tau)~\cite{Aubert:2007mh}\\
{\cal{B}}(\tau^- \to  K^-   K^-   K^+ \nu_\tau)~\cite{Aubert:2007mh}\\
{\cal{B}}(\tau^- \to  3h^- 2h^+ \nu_\tau ~(\mathrm{ex.~}K^0))~\cite{Aubert:2005waa}\\
{\cal{B}}(\tau^- \to  2\pi^- \pi^+ \eta \nu_\tau ~(\mathrm{ex.~}K^0))~\cite{Aubert:2008nj}
\end{eqnarray*}
\item and 10 measurements from Belle:
\begin{eqnarray*}
{\cal{B}}(\tau^- \to h^- \pi^0 \nu_\tau)~\cite{Fujikawa:2008ma}\\
{\cal{B}}(\tau^- \to \bar{K}^0 \pi^- \nu_\tau)~\cite{Epifanov:2007rf}\\
{\cal{B}}(\tau^- \to \pi^- \pi^- \pi^+ \nu_\tau ~(\mathrm{ex.~}K^0))~\cite{Lee:2010tc}\\
{\cal{B}}(\tau^- \to  K^-  \pi^- \pi^+ \nu_\tau ~(\mathrm{ex.~}K^0))~\cite{Lee:2010tc}\\
{\cal{B}}(\tau^- \to  K^-  \pi^-  K^+ \nu_\tau)~\cite{Lee:2010tc}\\
{\cal{B}}(\tau^- \to  K^-   K^-   K^+ \nu_\tau)~\cite{Lee:2010tc}\\
{\cal{B}}(\tau^- \to  \pi^- \pi^0 \eta \nu_\tau)~\cite{Inami:2008ar}\\
{\cal{B}}(\tau^- \to  K^- \eta \nu_\tau)~\cite{Inami:2008ar}\\
{\cal{B}}(\tau^- \to  K^- \pi^0 \eta \nu_\tau)~\cite{Inami:2008ar}\\
{\cal{B}}(\tau^- \to  \bar{K}^0 \pi^- \eta \nu_\tau)~\cite{Inami:2008ar}
\end{eqnarray*}
\end{itemize}

All of these 153 measurements are expressed as a linear function
of the form $(\sum_i \alpha_i P_i )\over(\sum_j \beta_j P_j )$ 
of ``base'' branching fractions $(P_i)$, 
which are chosen such that they sum up to unity.
The results of the fit are shown in Table~\ref{tab:TauGlobalFit},
which has $\chi^2/\mathrm{ndof} = 143.4/117$ (CL = 4.9\%). 

\begin{figure}[!hbtp]
\begin{center}
\includegraphics[height=.4\textheight,width=.49\textwidth]{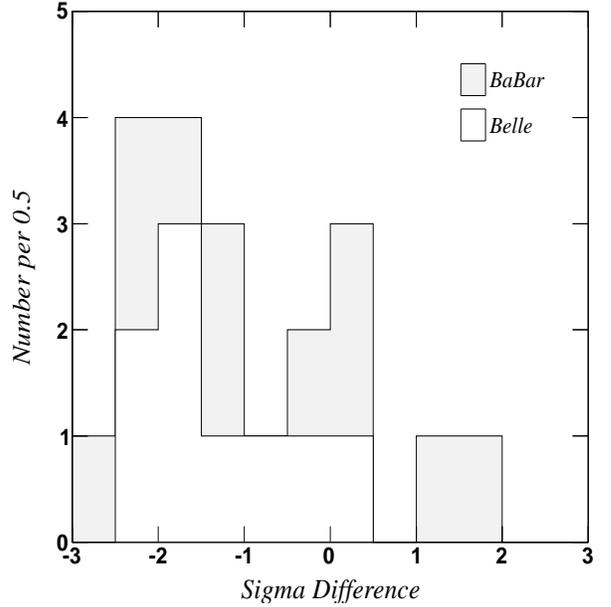}
\end{center}
\caption{Normalized difference of B-factory measurements w.r.t average of non-B-factory measurements.}
\label{fig:comp_NoBB}
\end{figure}
 
The differences between the global fit performed by PDG and us are:
\begin{itemize}
\item We expand the list of base modes from 31 to 37 to account for recent
  measurements by the B-factories with significant precision.
\item We use the original correlation matrix published by
 ALEPH~\cite{Schael:2005am} between hadronic modes instead of between
 pionic modes.
\item We use updated background estimates to adjust the B-factory measurements.
\item We avoid applying the PDG-style scale factors to all our measurements.
However, the BaBar and Belle measurements of ${\cal{B}}({\tau^- \to
  K^-K^+K^-\nu_\tau})$ are $5.4~\sigma$ apart. We scale the errors on
these 2 measurements only, following the PDG procedure for
single-quantity averaging.
\end{itemize}

\begin{table}[!hbtp]
\begin{tabular}{l| c} \hline
Base       modes                                                    & Branching fractions \\ 
(from $\tau^-$ decay)                                               & (in \%)\\ \hline
\multicolumn{2}{c}{leptonic modes} \\ \hline
$e^- \bar{\nu}_e \nu_\tau$                                            & 17.833 $\pm$  0.040 \\ 
$\mu^- \bar{\nu}_\mu \nu_\tau$                                        & 17.408 $\pm$  0.038 \\ 
\hline 
\multicolumn{2}{c}{non-strange modes} \\ \hline
$\pi^- \nu_\tau$                                                      &10.831 $\pm$  0.051 \\
$\pi^- \pi^0 \nu_\tau$                                                &25.531 $\pm$  0.090 \\
$\pi^- 2\pi^0 \nu_\tau ~(\mathrm{ex.~}K^0)$                           &9.278 $\pm$ 0.097 \\ 
$\pi^- 3\pi^0 \nu_\tau ~(\mathrm{ex.~}K^0)$                           & 1.046 $\pm$ 0.074 \\ 
$h^- 4\pi^0 \nu_\tau ~(\mathrm{ex.~}K^0,\eta)$                        & 0.107 $\pm$ 0.039 \\ 
$K^- K^0 \nu_\tau$                                                    & 0.160 $\pm$ 0.016 \\ 
$K^- \pi^0 K^0 \nu_\tau$                                              & 0.162 $\pm$ 0.019 \\ 
$\pi^- K_S^0 K_S^0 \nu_\tau$                                          & 0.024 $\pm$ 0.005 \\ 
$\pi^- K_S^0 K_L^0 \nu_\tau$                                          & 0.119 $\pm$ 0.024 \\ 
$\pi^- \pi^- \pi^+ \nu_\tau ~(\mathrm{ex.~}K^0,\omega)$               & 8.983 $\pm$ 0.050 \\ 
$\pi^- \pi^- \pi^+ \pi^0 \nu_\tau ~(\mathrm{ex.~}K^0,\omega)$         & 2.751 $\pm$ 0.069 \\ 
$h^- h^- h^+ 2\pi^0 \nu_\tau ~(\mathrm{ex.~}K^0,\omega,\eta)$         & 0.097 $\pm$ 0.036 \\ 
$h^- h^- h^+ 3\pi^0 \nu_\tau$                                         & 0.032 $\pm$ 0.003 \\ 
$\pi^- K^- K^+ \nu_\tau$                                              & 0.144 $\pm$ 0.003 \\ 
$\pi^- K^- K^+ \pi^0 \nu_\tau$                                        & 0.006 $\pm$ 0.002 \\ 
$3h^- 2h^+ \nu_\tau ~(\mathrm{ex.~}K^0)$                              & 0.082 $\pm$ 0.003 \\ 
$3h^- 2h^+ \pi^0 \nu_\tau ~(\mathrm{ex.~}K^0)$                        & 0.020 $\pm$ 0.002 \\ 
$\pi^- \pi^0 \eta \nu_\tau$                                          & 0.139 $\pm$ 0.007 \\ 
$\pi^- \omega \nu_\tau$                                              & 1.959 $\pm$ 0.064 \\ 
$h^- \pi^0 \omega \nu_\tau$                                          & 0.409 $\pm$ 0.042 \\ 
\hline
\multicolumn{2}{c}{strange modes} \\ \hline
$K^- \nu_\tau$                                                       & 0.697 $\pm$ 0.010 \\ 
$K^- \pi^0 \nu_\tau$                                                 & 0.431 $\pm$ 0.015 \\ 
$K^- 2\pi^0 \nu_\tau ~(\mathrm{ex.~}K^0)$                            & 0.060 $\pm$ 0.022 \\ 
$K^- 3\pi^0 \nu_\tau ~(\mathrm{ex.~}K^0,\eta)$                       & 0.039 $\pm$ 0.022 \\ 
$\bar{K}^0 \pi^- \nu_\tau$                                          & 0.831 $\pm$ 0.018 \\ 
$\bar{K}^0 \pi^- \pi^0 \nu_\tau$                                    & 0.350 $\pm$ 0.015 \\ 
$\bar{K}^0 \pi^- 2\pi^0 \nu_\tau$                                   & 0.035 $\pm$ 0.023 \\ 
$\bar{K}^0 h^- h^- h^+ \nu_\tau$                                     & 0.028 $\pm$ 0.020 \\ 
$K^- \pi^- \pi^+ \nu_\tau ~(\mathrm{ex.~}K^0,\omega)$                & 0.293 $\pm$ 0.007 \\ 
$K^- \pi^- \pi^+ \pi^0 \nu_\tau ~(\mathrm{ex.~}K^0,\omega,\eta)$     & 0.041 $\pm$ 0.014 \\  
$K^- \phi \nu_\tau (\phi \to KK)$                                   & 0.004 $\pm$ 0.001 \\ 
$K^- \eta \nu_\tau$                                                 & 0.016 $\pm$ 0.001 \\ 
$K^- \pi^0 \eta \nu_\tau$                                           & 0.005 $\pm$ 0.001 \\ 
$\bar{K}^0 \pi^- \eta \nu_\tau$                                     & 0.009 $\pm$ 0.001 \\ 
$K^- \omega \nu_\tau$                                              & 0.041 $\pm$ 0.009 \\ 
\hline
Sum of strange modes                                              & 2.8796 $\pm$ 0.0501 \\ \hline
Sum of all modes                                                  & 100.00              \\ \hline
\end{tabular}\\[2pt]
\caption{Results of unitarity constrained fit.}
\label{tab:TauGlobalFit}
\end{table}

The three-kaon decays had not been observed before the B-factory era.
For the remaining 20 B-factory measurements, 
Figure~\ref{fig:comp_NoBB} shows a histogram of the normalized difference 
((B-factory value minus averaged non-B-factory value)/estimated uncertainty in the difference).
The average difference between the two sets of measurements is -0.98 $\sigma$ 
(-1.26 $\sigma$ for the 9 Belle measurements and -0.75 $\sigma$ for the 11 BaBar measurements).
Although this systematic trend is yet to be understood,
the magnitude of discrepancy is less than the value of reported by the PDG~\cite{PDG_2010}.

\section{Tests of Lepton Universality}
\label{sec:Tau_LU}

From the unitarity constrained fit, we obtain
$
{{\cal{B}}({\tau^- \to \mu^- \overline{\nu}_\mu \nu_\tau})}/
{{\cal{B}}({\tau^- \to e^- \overline{\nu}_e  \nu_\tau})}$
= 
$0.9762\, \pm\,   0.0028$,
which includes a correlation of $18.33\%$
between the branching fractions.
This yields a value of $\left( \frac{g_\mu}{g_e} \right)$ = 
$1.0019\, \pm\, 0.0014$,
which is consistent with the Standard Model (SM) value.

Using the world averaged mass, lifetime and meson decay rates~\cite{PDG_2010}, we determine 
$\left( \frac{g_{\tau}}{g_{\mu}} \right)$ =
$0.9966\, \pm\, 0.0030$ $(0.9860\, \pm\, 0.0073)$
from the pionic and kaonic branching fractions,
with a correlation of $13.10\%$.
Combining these results, we obtain 
$\left( \frac{g_{\tau}}{g_{\mu}} \right)$ = $0.9954\, \pm\, 0.0029$,
which is consistent with ($1.6~\sigma$ below) the SM expectation.

\begin{figure}[!hbtp]
\begin{center}
\begin{minipage}{.42\textwidth}
\includegraphics[height=.3\textheight]{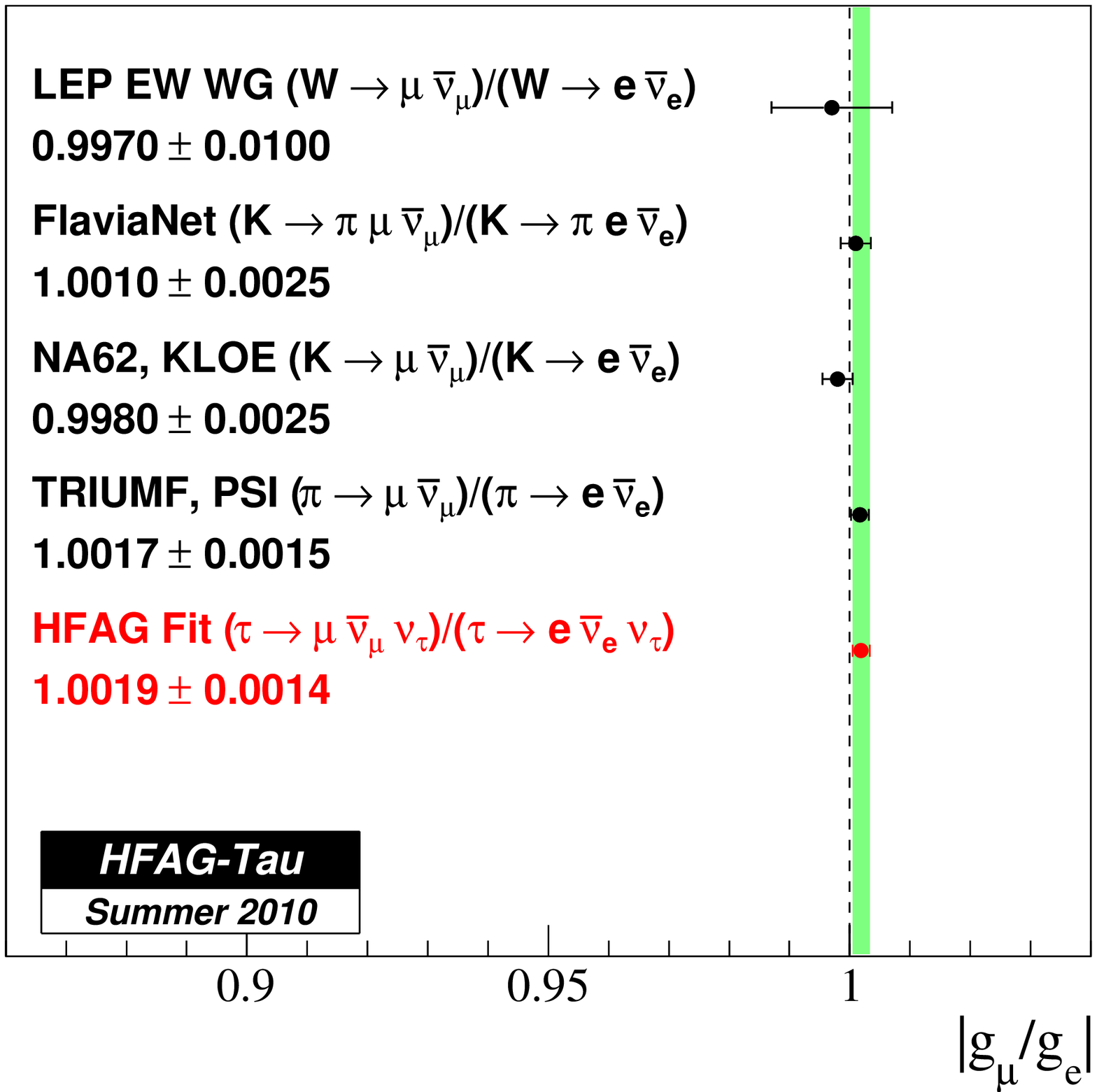}
\includegraphics[height=.3\textheight]{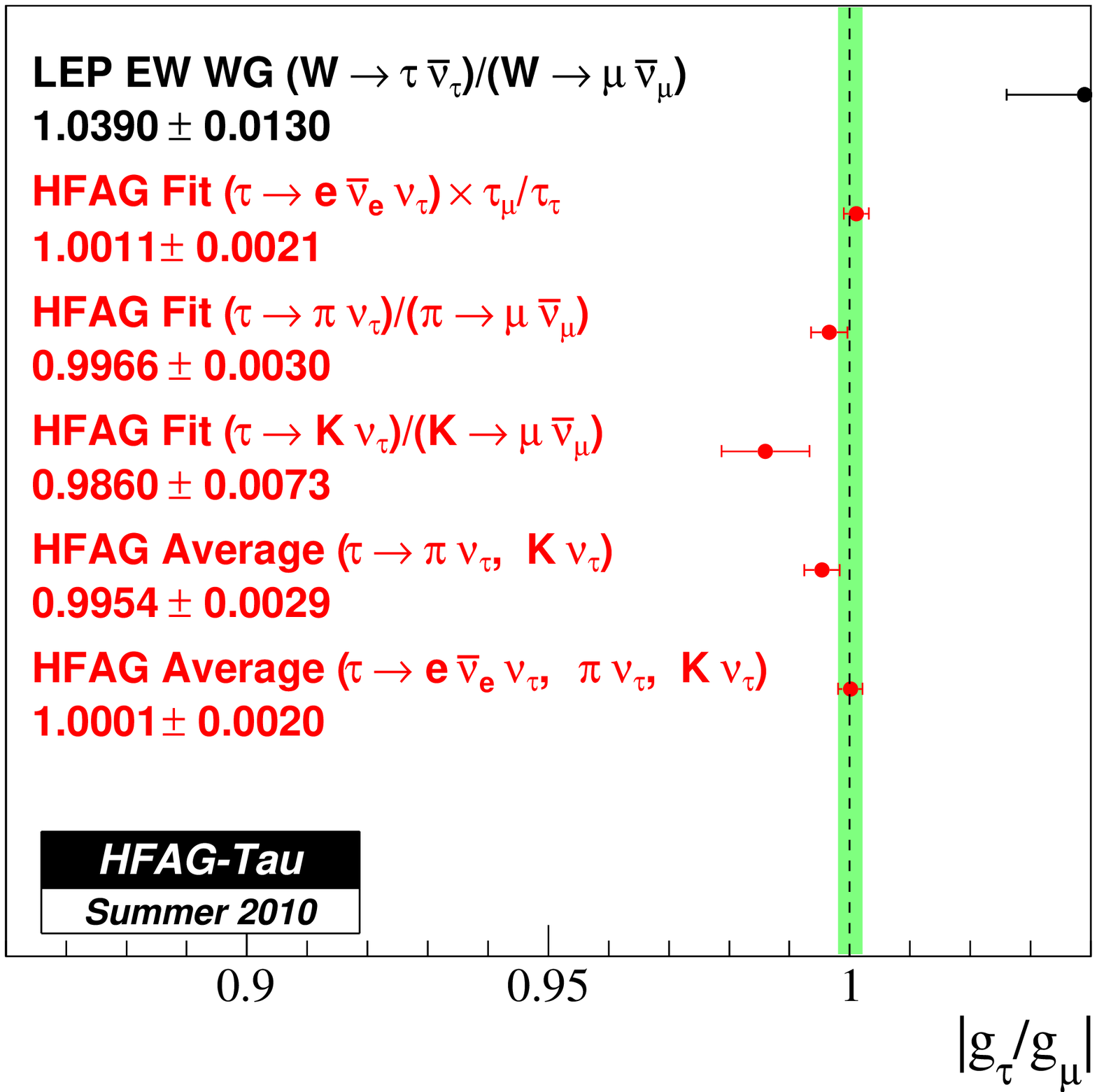}
\includegraphics[height=.3\textheight]{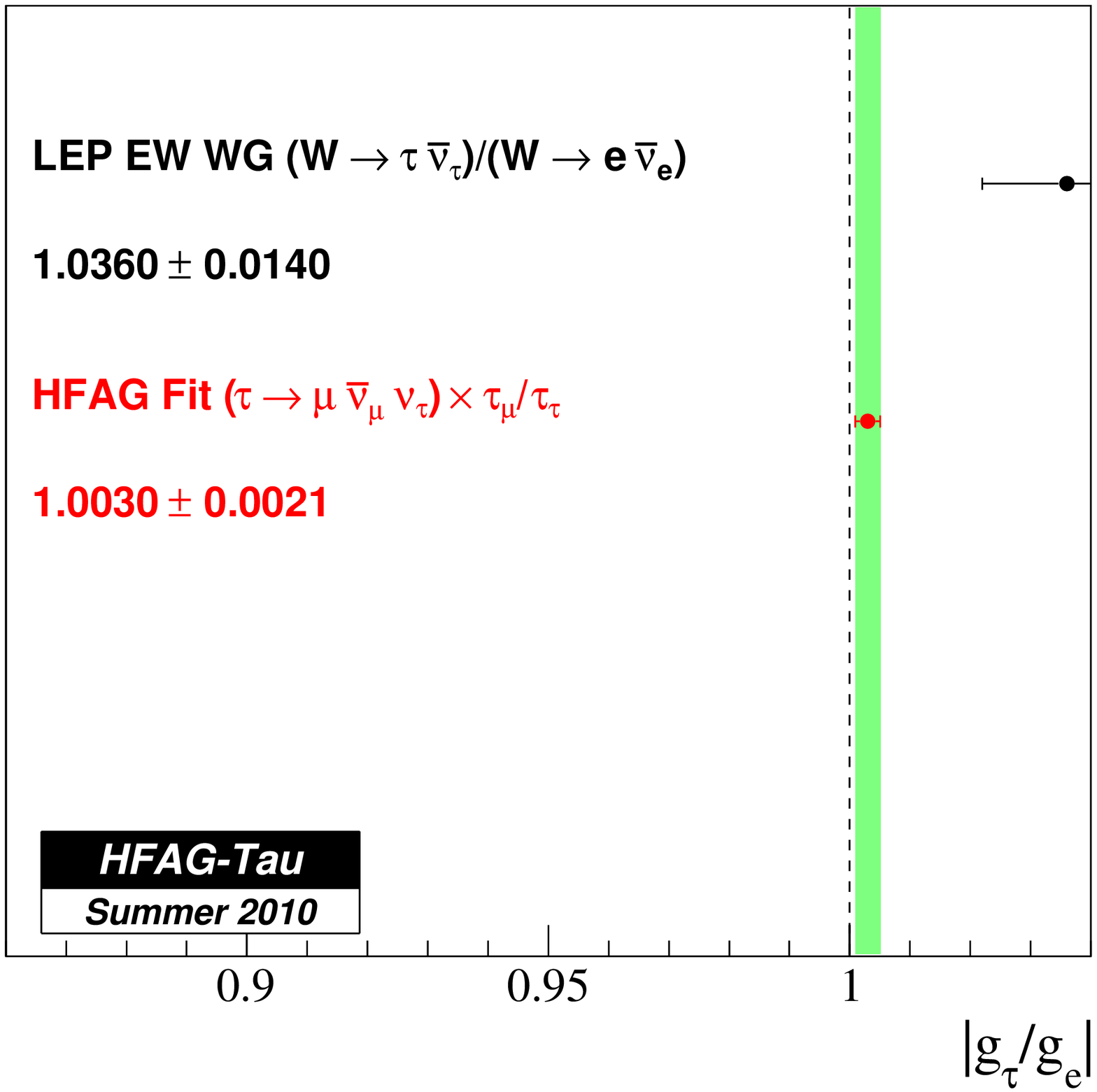}
\end{minipage}
\end{center}
\caption{Measurements of lepton universality from W, kaon, pion and tau decays.}
\label{fig:TauLU}
\end{figure}

We also test lepton universality between $\tau$ and $\mu$ ($e$),
by comparing the averaged electronic (muonic) branching fractions of
the $\tau$ lepton with the predicted branching fractions from measurements
of the $\tau$ and $\mu$ lifetimes and their respective masses~\cite{PDG_2010}.
This gives 
$\left( \frac{g_{\tau}}{g_{\mu}} \right)$ = $1.0011\, \pm\, 0.0021$
and
$\left( \frac{g_{\tau}}{g_{e}} \right)$ = $1.0030\, \pm\, 0.0021$.
The correlation co-efficient between the determination of
$\left( \frac{g_{\tau}}{g_{\mu}} \right)$ from electronic branching
fraction with the ones obtained from pionic and kaonic branching fractions
are $48.16\%$ and  $21.82\%$, respectively.
Averaging the three values, we obtain
$\left( \frac{g_{\tau}}{g_{\mu}} \right)$ = $1.0001\, \pm\, 0.0020$,
which is consistent with the SM value.
In Figure~\ref{fig:TauLU}, we compare these determinations 
with the values obtained from pion~\cite{PiToMu},
kaon~\cite{Antonelli:2010yf} and W decays~\cite{Alcaraz:2006mx}.

\section{Measurement of $|V_{us}|$}
\label{sec:Tau_Vus}

\newcommand{\Vudvalue} {\ensuremath{0.97425\, \pm\, 0.00022}\xspace}

Using the unitarity constraint on the first row of the CKM matrix
and $\Vud = \Vudvalue$~\cite{Hardy:2008gy}, one gets $\Vus = 0.2255\, \pm\, 0.0010$.
Here we present 3 extractions for $|V_{us}|$ using
${\cal{B}}({\tau^-\to K^-\nu_\tau})$, 
${{\cal{B}}({\tau^- \to K^-\nu_\tau})}/
{{\cal{B}}({\tau^- \to \pi^-\nu_\tau})}$,
and inclusive sum of $\tau$ branching fractions having net strangeness of unity in the final state.

We use the value of kaon decay constant $f_K = 157\, \pm\, 2 \mev$ 
obtained from Lattice calculations~\cite{Follana:2007uv}, 
and our value of ${\cal{B}}({\tau^-\to K^-\nu_\tau}) =$
$\frac{G^2_F f^2_K \Vus^2 m^3_{\tau} \tau_{\tau}}{16\pi\hbar} \left (1 - \frac{m_K^2}{m_\tau^2} \right )^2 S_{EW},$
where $S_{EW} = 1.0201\, \pm\, 0.0003$~\cite{Erler:2002mv},
to determine $\Vus= 0.2204\, \pm\, 0.0032$,
which is consistent with ($1.5~\sigma$ below) the value from CKM unitarity.

We extract $\Vus$ using $\Vud$, Lattice calculation of the ratio of decay constants
$f_K/f_\pi = 1.189\, \pm\, 0.007$~\cite{Follana:2007uv} and
the long-distance correction $\delta_{LD} = (0.03\, \pm\, 0.44)\%$,  
estimated using corrections to $\tau\to h\nu_\tau$ and $h \to \mu\nu_\mu$~\cite{LongDistance},
for the ratio
$$\frac{{\cal{B}}({\tau^- \to K^-\nu_\tau})}{{\cal{B}}({\tau^- \to \pi^-\nu_\tau})} 
=
\frac{f_K^2 |V_{us}|^2}{f_\pi^2 |V_{ud}|^2}
 \frac{\left( 1 -  \frac{m_K^2}{m_\tau^2} \right)^2}{\left( 1 -  \frac{m_\pi^2}{m_\tau^2} \right)^2} (1+\delta_{LD}),$$
where short-distance electro-weak corrections cancel in this ratio
measured to be
${{\cal{B}}({\tau^- \to K^-\nu_\tau})}/{{\cal{B}}({\tau^- \to \pi^-\nu_\tau})}$
$=$ $ 0.0644\, \pm\, 0.0009$.
This includes a correlation of $-0.49\%$
between the branching fractions, and yields $\Vus = 0.2238\, \pm\, 0.0022$,
which is also consistent with ($0.7~\sigma$ below) $\Vus$ from CKM unitarity.
 
The total hadronic width of the \mtau normalized to the electronic branching fraction,
$R_{\rm{had}} = {\cal{B}}_{\rm{had}}/{\cal{B}}_{\rm{e}}$,
can be written as $R_{\rm{had}} = R_{\rm{non-strange}} + R_{\rm{strange}}$.
We can then measure
$$|V_{us}| = \sqrt{R_{\rm{strange}}/\left[\frac{R_{\rm{non-strange}}}{|V_{ud}|^2} -  \delta R_{\rm{theory}}\right]}.$$

Here, we use $\Vud = \Vudvalue$~\cite{Hardy:2008gy}, and $\delta R_{\rm{theory}} = 0.240\, \pm\, 0.032$~\cite{Gamiz:2006xx} 
which contributes to an error of $0.0010$ on $|V_{us}|$.
We note that this error is equivalent to  half the difference between calculations of $|V_{us}|$ 
obtained using fixed order perturbation theory (FOPT) and 
contour improved perturbation theory (CIPT) calculations of $\delta R_{\rm{theory}}$~\cite{Maltman:2010hb}, 
and twice as large as the theoretical error proposed in Ref.~\cite{Gamiz:2007qs}.

As in Ref.~\cite{Davier:2005xq}, we improve upon the estimate of
electronic branching fraction by averaging its direct measurement 
with its estimates of $(17.899\, \pm\, 0.040)\%$ and $(17.794\, \pm\, 0.062)\%$
obtained from the averaged values of muonic branching fractions and 
the averaged value of the lifetime of the 
\mtau lepton = $(290.6\, \pm\, 1.0) \times 10^{-15} ~\mathrm{s}$~\cite{PDG_2010},
assuming lepton universality and taking into account the correlation between the leptonic branching fractions.
This gives a more precise estimate for the electronic branching fraction: ${\cal{B}}^{\rm{uni}}_{\rm{e}}$ = $ (17.852\, \pm\, 0.027)\%$.

Assuming lepton universality, the total hadronic branching fraction 
can be written as: ${\cal{B}}_{\rm{had}} = 1 - 1.972558 ~ {\cal{B}}^{\rm{uni}}_{\rm{e}}$,
which gives a value for the total \mtau hadronic width normalized to 
the electronic branching fraction as $R_{\rm{had}} = 3.6291\, \pm\, 0.0086$.

Non-strange width is $R_{\rm{non-strange}} = R_{\rm{had}} - R_{\rm{strange}}$,
where the value for the strange width $R_{\rm{strange}}$ = $0.1613\, \pm\, 0.0028$
is obtained from the sum of the strange branching fractions 
with the unitarity constrained fit as listed in Table~\ref{tab:TauGlobalFit}.
This gives a value of  $\Vus$ $=$ $0.2174\, \pm\, 0.0022$,
which is $3.3~\sigma$ lower than the CKM unitarity prediction.


\begin{figure}[!hbtp]
\begin{center}
\includegraphics[height=.38\textheight,width=.49\textwidth]{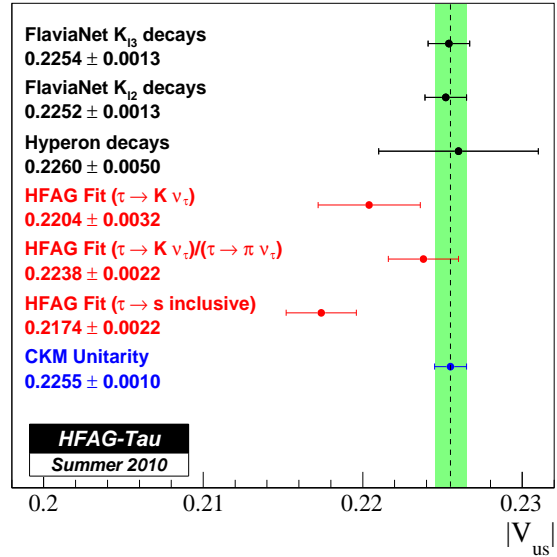}
\end{center}
\caption{Measurements of $\Vus$ from kaon, hyperon and tau decays.}
\label{fig:Vus}
\end{figure}
 
Summary of these $\Vus$ values are plotted in Figure~\ref{fig:Vus}, 
where we also include values from hyperon and kaon decays~\cite{Antonelli:2010yf}.

\section{Search for lepton flavor violation in $\tau$ decays}
\label{sec:Tau_LFV}

\begin{figure*}[!hbtp]
\begin{center}
\includegraphics[height=.4\textheight]{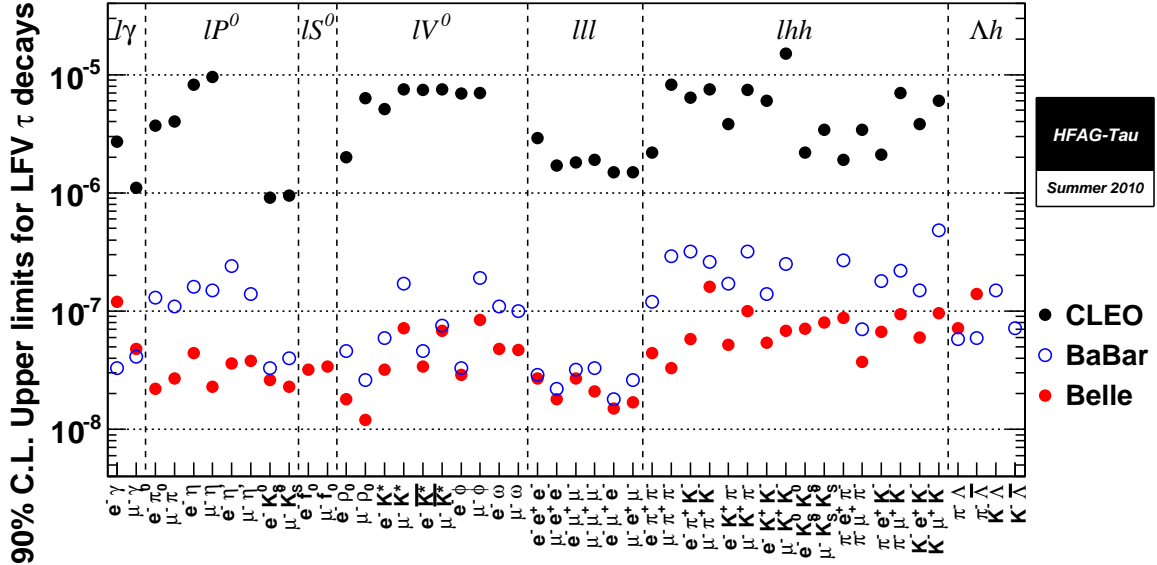}
\end{center}
\caption{Status of searches for lepton flavor violation in $\tau$ decays.}
\label{fig:Tau_LFV}
\end{figure*} 

The status of searches for lepton flavor violation in $\tau$ decays is 
summarized in Figure~\ref{fig:Tau_LFV}. A table of these results and 
the corresponding references are provided on the HFAG-Tau web site.

\end{document}